\documentstyle[12pt,epsf]{article}
\begin{document}

\sloppy

\title{The Pion Photoproduction in the $\Delta$(1232) Region}

\author{Chung Wen Kao and Thomas D. Cohen \\
Department of Physics, University of Maryland\\
College Park, MD 20742-4111\\
\footnotesize DOE/ER/40762-166~~~~UMD PP\#99-042}

\maketitle

\rm
  
\begin{abstract}

We investigate the pion photoproduction off the nucleon in the
$\Delta$ region in the framework of an effective chiral Lagrangian
including pions, nucleon, and $\Delta(1232)$.  We work to third order
in a small scale expansion with both $m_{\pi}$ and $M_{\Delta} -M_{N}$
treated as light scales.  We note that in the $\Delta$ region,
straightforward power counting breaks down as the amplitude becomes
very large.  To deal with this problem, we suggest that the
appropriate way to compare the theoretical calculations with
experimental data is via weighted integrals of the amplitudes through
the $\Delta$ region. 
\end{abstract}

\pagebreak
\section{ Introduction}

The $\Delta$(1232) resonance enjoys a special status in the family of
the nucleon resonances. It lies only about 300 MeV above the nucleon
ground state, nicely isolated from the plethora of other densely
populated resonances at higher energies, and only strongly couples to the
$N\pi$ system. Therefore, the matrix of physical current-like $\langle
N|J^{em}_{\mu}|\Delta \rangle$ is supposed to be easily extracted from
experimental observables and expected to shed light on their
structures.

 For example, in simple constituent quark models, the interaction
between quarks is believed to have a tensor part due to the color
hyperfine interaction\cite{RGD}. One consequence is the admixture of
higher orbital angular momentum components in s-state quark wave
functions of nucleon ground state and $\Delta$.  The d-state mixture
allows for an electric quadrupole E2 transition in
$\gamma N\rightarrow \Delta(1232)$ excitation, which is otherwise 
a pure magnetic dipole M1 transition.
The M1 and E2 transitions can be directly excited by photons and the 
subsequent 
pion decay can be observed. The amplitudes in $\pi$N final states are
usually denoted by $E^{I}_{l_{\pm}}$ and $M^{I}_{l_{\pm}}$ where E and
M are respectively electric and magnetic multipoles, $l$ is the orbital
angular momentum of the pion, the $\pm$ signs refer to the total angular
momentum $j=l\pm 1/2$, and I is the isospin of the $\pi$N system.  Therefore,
in this class of models, the ratio $R_{EM}\equiv
E2/M1=E^{3/2}_{1+}/M^{3/2}_{1+}$ is related to the tensor component in
the interaction of quarks.  The simplest noninteracting SU(6) quark
model predicts $R_{EM}$=0;
the nonrelativistic constituent 
quark model with gluon exchange predicts a rather small and 
negative result ranging between $-0.008\%$ to $-2\%$\cite{IKK},\cite{IKK2}.
  
Many baryon models have the pion cloud playing an important role.  Because of the derivative coupling
pions are relatively efficient at generating strength for the E2. 
Cloudy bag models lead to $-2\%<R_{EM}<0$\cite{WK}. Larger negative
values are given by Skyrme and other hedgehog models:$
-5.9\%<R_{EM}<-2.5\%$\cite{WW,CB0}, and exchange current yields:
$R_{EM}=-3.5\%$\cite{BHF}.  Thus $R_{EM}$ becomes a sensitive test for
different models for baryon structure.

On the other hand, for several different reasons it is important to
understand the $N\rightarrow \Delta$ EM transition and more generally,
the nucleon and $\Delta$-isobars, without reference to any particular
baryon model.  For example, to study the photoproduction of the mesons
off complex nuclei\cite{A1}, it is necessary to use single nucleon
information such as $\gamma \Delta N$ vertex strength as input.  Such
investigations help to clarify how the effective degrees of freedom,
nucleons, mesons, and $\Delta$-isobars play their roles in nuclei.
$\Delta$-isobars by themselves are particularly interesting objects in
the large $N_{c}$ QCD because they will be degenerate with the nucleon
if $N_{c}$ exactly goes to infinity.  By a simple argument based on
$N_{c}$ counting rules applied to the $\pi N$ scattering, it was shown
that large $N_{c}$ QCD with only I=J=1/2 nucleon interacting with
pions is inconsistent \cite{DJM}; other states in the tower of
I=J=1/2,3/2 ... must be included to satisfy the consistency
conditions. Therefore, the study of $N\rightarrow \Delta$ transitions
may help provide insights into the large $N_{c}$ QCD.  Moreover,
$R_{EM}$ was predicted to be unity in the domain of perturbative
QCD\cite{E}; thus the study of $N\rightarrow \Delta$ EM transition may
provide a window into the breakdown of pQCD as momentum transfer
drops.
    
It is difficult to determine the resonant $E^{3/2}_{1+}$ for two
reasons.  From the experimental side, the precise measurement of this
quadrupole amplitude is difficult due to its smallness compared to
dominant magnetic dipole amplitude.  Recently there has been
substantial progress so that precise measurements of spin observables
are now possible.  At Mainz\cite{B} a $p(\vec{\gamma}p)\pi^{0}$
measurement was performed with tagged linely polarized photons
produced at 855 MeV Mainz Microtron MAMI; the differential cross
section and photon asymmetry $\Sigma$ were measured simultaneously for
the $p\pi^{0}$ and $n\pi^{+}$ channels. They took
$R_{EM}=ImE_{1+}^{3/2}/ImM_{1+}^{3/2}$ at the peak of the
$\Delta(1232)$ resonance, and the reported value is $R_{EM}=-(2.5\pm
0.2\pm 0.2)\%$. There are also analyses of other groups
based on their and 
BNL's data\cite{KS}--\cite{W}.

However, there is a more serious problem on the theoretical side: The
non-resonant contribution in $E^{3/2}_{1+}$ cannot be separated
directly by the measurement\cite{WWA2}.  Thus, the direct comparison
between the experimental $R_{EM}$ with the results of the calculations
based on the specific baryon models are meaningless unless one extends
the model to include continuum states in a consistent way.  Various
methods have been proposed for the isolation of the resonant part from
the measured multipoles results but their answers vary greatly.
Different prescriptions result in different kinds of definitions of
the resonant E2/M1 ratio\cite{WWA}.
 
For example, in the effective Lagrangian approach one main difficulty
of this separation is due to the unitarization. The Born terms and the
$\Delta$ excitation treated as the leading tree graphs, do not fulfill
the requirement of unitarity\cite{w}. The unitarity may be put in by hand
but different unitarization prescriptions lead to different separations
between the resonant part and background 
contributions\cite{RMW}--\cite{NO}. 

On the other hand, models \cite{SNY},\cite{TO}.  which
treat pion photoproduction dynamically (i.e. solving the
corresponding Lippman-Schwinger equation for a given $\pi N$
interaction) are automatically unitary since the rescattering process
is included. Such models also provide a basis for the analysis of the
role of final state interactions.  However, to solve the dynamical
equation, some phenomenological form factors must be included.  They 
are needed to regularize
the driving terms of the interaction. Since these form factors are put
in by hand, they also make the separation between the resonant part and
background contributions model dependent\cite{SL}.

Recently, the ``speed plot technique'' was also applied to this
problem\cite{HDT}. The ``speed'' SP of the scattering amplitude T is
defined by: 
$$SP(W)=|\frac{dT(W)}{dW}|.$$ 
Here W is the total c.m
energy. In the vicinity of the resonance pole the energy dependence of
the full amplitude $T=T_{R}+T_{B}$ is determined by the resonance
contribution:
$$T_{R}(W)=\frac{r\Gamma_{R}e^{i\phi}}{M_{R}-W-i\Gamma_{R}/2},$$ 
while $T_{B}$ is a smooth function of energy. Application of this method to
the amplitude of Tiator et al. derived by fixing the $t$ dispersion
relation gives 
$$R_{\Delta}\equiv \frac{r_{E}e^{i\phi_{E}}} 
{r_{M}e^{i\phi_{M}}}=-0.035-0.046i.$$ 
Th. Wilbois et al.\cite{WWA}
suggested a more sophisticated way to implement this and found
$R_{\Delta}=-0.040-0.047i$. Note this $R_{\Delta}$ is a complex and
energy-independent quantity; unfortunately it remains unclear how
$R_{\Delta}$ could be compared to any microscopic baryon model, and
it is difficult to determine what, if anything, about the chiral
dynamics of $\pi\Delta N$ system.

Some of the ambiguities mentioned above will be ameliorated by using
an approach based on chiral perturbation theory.  Unitarity is
guaranteed, at least perturbatively, because loop graphs are
included. Since dimensional regularization can be implemented
straightforwardly in this approach, one avoids the model dependence
inherent in the introduction of phenomenological form factors.
Therefore ChPT provides a model-independent picture of pion
photoproduction ---at least to the extent the expansion converges.
Many previous ChPT calculations were limited to the threshold region
because $\Delta$-isobars were not included explicitly.  In such
calculations, the $\Delta$ effects are supposed to be included in the
form of local counterterms.  The essential physical idea is that the
delta propagator is treated as though it was shrunken to a point and
the energy dependence of $\Delta$-isobars are reproduced by higher 
dimension operators which are suppressed by 
$1/(M_{\Delta}-M_{N})$\cite{BKM1},\cite{BKM2}.  
Of course, such a framework cannot be used for our problem since we wish to
work in the $\Delta$ region.

It has been recognized, even in the threshold region, that the
$\Delta$ is a low energy excitation and it is presumably sensible to
include it dynamically, creating a more general effective field theory
than simple ChPT.  This is the spirit of the original work of Jenkins
and Manohar\cite{JM1},\cite{JM2}, and the applications by Butler,
Savage and Springer to the SU(3)$\times$SU(3) case\cite{BSS1},\cite{BSS2} 
although their works are never beyond the leading order. 

 Recently,
Hermmert et al.\cite{HHK}--\cite{k} have formalized such an approach.
They have developed a consistent power expansion scheme, the so-called
``small scale expansion'', which allows for nucleon and
$\Delta$-isobar degrees of freedom to be treated simultaneously in an
SU(2) effective chiral Lagrangian.  Whereas in conventional HBChPT one
expands in power of external momenta in analogy to the meson sector; a
phenomenological expansion was set up in the small scale
$\epsilon$. ``Small scale'' denotes the soft momentum, the pion mass
or the mass difference $\Delta\equiv M_{\Delta}-M_{N}$.  One natural
reason to do this is: $\Delta$ now is treated as a new dimensionful
parameter which stays finite in chiral limit but is nevertheless of
comparable size to $m_ \pi$ in the real world. To assert the accuracy
of this novel approach one has to systematically calculate observables
and compare the resulting predictions. The $\gamma N\rightarrow
\Delta$ transition seems to be a promising case to implement such a
scheme.\cite{GHKP}

There is a potential problem, however.  The perturbative power
counting scheme fails in producing calculations for momenta transfers
in the $\Delta$ region.  This is unfortunate since this is precisely the
region where we wish to work.  The reason for this failure is quite
obvious.  The generic power counting has the $\Delta$ propagator
behaving as ${\cal O}(1/{\epsilon})$, where $\epsilon$ is the small
dimensionless parameter, $\epsilon \equiv m_\pi/\Lambda,
\Delta/\Lambda$.  For generic low momenta this is valid.  As one
approaches the $\Delta$ resonance, the propagator, treated at lowest
order diverges, spoiling the power counting.  One may hope to cure
this by including the $\Delta$ self-energy in the propagator.  The
imaginary part of the self energy remains finite as there is a
physical decay channel. There is a difficulty with this approach;
namely, certain graphs are iterated to all orders---the $\Delta$ self
energy insertions---while others are not.  This makes it very difficult
to assure that one has a systematic power counting scheme.  However,
even if the propagator is anomalously large in the neighborhood of the
resonance, we note that integrals of the propagator times smooth
functions over regions of order $\epsilon \Lambda$ are order
$1/\epsilon$---the same as the generic $\Delta$ propagator.  Clearly,
this means that so long as we are only sensitive to integrals over the
propagator the power counting is safe.  Loop diagrams with $\Delta$'s
as intermediate states are of this type.  However, there are also
contributions coming from graphs such as the Born graphs, in which the
four momentum flowing through the propagator is fixed by the external
momenta.  When those external momenta are such that four momentum in
the $\Delta$ propagator are close to the pole the power counting has
broken down.

We propose a possible way to avoid this problem.  The power counting
scheme used simply does not allow us to accurately calculate the
amplitudes in the vicinity of the $\Delta$ pole.  However, if we limit
our ambitions to asking questions about integrals of the amplitudes
through the $\Delta$ region the power counting scheme remains viable.
Accordingly, our proposal is that one should not directly compare
calculated amplitudes with experiment.  Rather one should extract the
amplitudes from experiment, estimate weighted integrals over the
amplitudes and compare these integrals of the experimental amplitudes
with theoretical ones.  In this way, we can make predictions of
quantities in the $\Delta$ region---albeit integrated quantities---for
which the power counting scheme is viable.  We note that this approach
has some important limitations.  The most obvious one, of course is
that we cannot make a direct prediction of the experimental
observables.  There is also an important practical limitation.  We do
not make predictions for integrals of differential cross sections but
for integrals of amplitudes.  In order to do this, one must fix the
amplitudes from the measurements.  Unfortunately, the various
spin-dependent differential cross sections each depend on several
amplitudes.  One needs to make several independent measurements to
extract the amplitudes\cite{CT}.  To the best of our knowledge none of
the amplitudes has been extracted from the experimental quantities to
date.  This means we can not presently use the methods discussed here
to compare with experiment.  However, future spin-dependent
measurements could alter this situation.

This paper is organized as following: In Sec. 2 the formalism of
HBChPT and the steps to include the $\Delta$ degree of freedom are
briefly sketched. Sections 3 and 4 describe the loop and Born
graphs.
Renormalization is discussed in section 5. Finally, in Sec. 6, the
method for comparing the theoretical calculation with experimental
data is reported; some related issues and further prospects are also
discussed.

\section{ Effective Lagrangian}

The starting point for our approach is the most general chiral
invariant Lagrangian involving relativistic spin 1/2 and spin 3/2
fields: 
\begin{equation} {\cal L}={\cal L}_{N}+{\cal
L}_{\Delta}+({\cal L}_{\Delta N}+h.c.).
\end{equation}
Although the expressions of the standard conventions of SU(2) HBChPT
exists in literature as in Ref. \cite{HHK2}, we believe that it is useful in
establishing our notation to give relevant expressions here: 
\begin{equation}
U=u^{2}=exp(\frac{i}{F_{\pi}}\vec{\tau}\cdot\vec{\pi}),
\end{equation}
\begin{equation}
\nabla_{\mu}U=\partial_{\mu}U-i(v_{\mu}+a_{\mu})U+iU(v_{\mu}-a_{\mu}),
\end{equation}
\begin{equation}
u_{\mu}=iu^{\dagger}\nabla_{\mu}Uu^{\dagger},
\end{equation}
\begin{equation}
w_{\mu}^{i}=\frac{1}{2}Tr(\tau^{i}u_{\mu}),
\end{equation}
\begin{equation}
w_{\mu\nu}^{i}=\frac{1}{2}Tr(\tau^{i}[D_{\mu},u_{\nu}]),
\end{equation}
\begin{equation}
\Gamma_{\mu}=\frac{1}{2}[u^{\dagger},\partial_{\mu}u]-\frac{i}{2}u^{\dagger}(v_{\mu}+a_{\mu})u-\frac{i}{2}u(v_{\mu}-a_{\mu})u^{\dagger},
\end{equation}
\begin{equation}
u^{ij}_{\mu}=\xi^{ik}u_{\mu}\xi^{kj}, \xi^{ij}=\delta^{ij}-\frac{1}{3}\tau^{i}\tau^{j},
\end{equation}
\begin{equation}
\Gamma_{\mu}^{ij}=\xi^{ik}\Gamma_{\mu}\xi^{kj},
\end{equation}
\begin{equation}
D_{\mu}^{ij}\psi_{j}^{\nu}=(\partial_{\mu}\delta^{ij}+\Gamma_{\mu}^{ij})\psi_{j}^{\nu},
\end{equation}
\begin{equation}
D_{\mu}\psi_{N}=\partial_{\mu} \psi_{N}+\Gamma_{\mu}\psi_{N},
\end{equation}
\begin{equation}
F_{\mu}^{R}=v_{\mu}+a_{\mu}, F_{\mu}^{L}=v_{\mu}-a_{\mu},
\end{equation}
\begin{equation}
F^{L,R}_{\mu\nu}=\partial_{\mu}F_{\nu}^{L,R}-\partial_{\nu}F_{\mu}^{L,R}-i[F^{L,R}_{\mu},F_{\nu}^{L,R}] ,
\end{equation}
\begin{equation}
f_{\mu\nu}^{\pm}=u^{\dagger}F^{R}_{\mu\nu}u\pm uF^{L}_{\mu\nu}u^{\dagger},
\end{equation}
\begin{equation}
f_{\pm}^{i\mu\nu}=\frac{Tr}{2}(\tau^{i}f_{\pm}^{\mu\nu}),
\end{equation}
\begin{equation}
f_{\pm}^{i\mu\nu ;\rho}=\frac{Tr}{2}(\tau^{i}[D^{\rho},f_{\pm}^{\mu\nu}]),
\end{equation}
\begin{equation}
\chi_{\pm}=u^{\dagger}\chi u^{\dagger}\pm u\chi^{\dagger}u,
\end{equation}
When the only external fields are photons, $f_{\mu\nu}^{\pm}$ simplifies to:
\begin{equation}
f^{\pm}_{\mu\nu}=e(\partial_{\mu}A_{\nu}-\partial_{\nu}A_{\mu})(uQu^{\dagger}\pm u^{\dagger}Qu),
\end{equation}
Q$\equiv$diag(1,0).
 $v_{\mu}$ denotes external vector field and $a_{\mu}$ denotes axial field.
Here $\chi(x)=s(x)+ip(x)$ includes the explicit chiral symmetry breaking 
through the small current quark masses, $s(x)$= B diag($m_{u},m_{d}$),
$B= \frac{|\langle 0| \bar{q}q|0\rangle|}{F_{\pi}}$, with $F_{\pi}$ 
the weak pion decay constant. Here isospin invariance ($m_{u}=m_{d}$) 
is assumed.

The spin-3/2 field is represented via a Rarita-Schwinger spinor\cite{RS}, i.e,
\ a vector-spinor field $\Psi_{\mu}(x)$ satisfies the equation of motion:   
\begin{equation}
(i\gamma_{\nu}\partial^{\nu}-M_{\Delta})\Psi_{\mu}(x)=0,
\end{equation}
with the subsidiary conditions:
\begin{equation}
\gamma_{\mu}\Psi^{\mu}(x)=0.
\end{equation}
This subsidiary condition ensures that the physical particle created
by the field is spin-3/2 as opposed to 1/2. 
Following Hermmert et al.\cite{HHK} the Lagrangian of $\Delta$-isobars 
is given by:
\begin{equation}
{\cal L}_{\Delta}=\bar{\psi^{\mu}_{i}} \Lambda_{\mu \nu}^{ij} \psi^{\nu}_{j},
\end{equation}
\begin{equation}
\begin{array}[c]{c}
\Lambda^{ij}_{\mu \nu}=-[(i\not\!D^{ij}-M_{\Delta}\delta^{ij})
g_{\mu \nu}-\frac{1}{4}\gamma_{\mu}\gamma^{\nu}
(i\not\!D^{ij}-M_{\Delta}\delta^{ij})\gamma_{\lambda}\gamma_{\nu}  \\ [.1in]
+\frac{g_{1}}{2}g_{\mu \nu}\not\!u^{ij}\gamma_{5}+\frac{g_{2}}{2}
(\gamma_{\mu}\not\!u^{ij}_{\nu}+\not\!u^{ij}_{\mu}\gamma_{\nu})\gamma_{5}
+\frac{g_{3}}{2}\gamma_{\mu}\not\!u^{ij}\gamma_{5}\gamma_{\nu}]  .\\
\end{array}
\end{equation}

${\cal L}_{N}$ is well known in HBChPT, to the third order it is 
given by\cite{EM},\cite{FMS}.
Only nine of the terms in ${\cal L}^{(3)}$ are relevant in our problem.
As Fettes et al.\cite{FMS} pointed out that there is one counterterm
too many in ${\cal L}^{(3)}$ in \cite{EM}, however the extra term
${\cal O}_{4}$ is irrelevant in the pion photoproduction process, so we still adopt the
basis of \cite{EM}.\\

The interaction of the $\Delta$ field with photons, pions, and nucleons is
given by the following effective Lagrangian\cite{PS},\cite{NL}:
\begin{equation}
{\cal L}_{\pi\Delta N}=
{\cal L}_{\Delta N}^{(1)}+{\cal L}_{\Delta N}^{(2)}+
{\cal L}_{\Delta N}^{(3)}, 
\end{equation}
\begin{equation}
{\cal L}_{\Delta N}^{(1)}=g_{\pi \Delta N}\bar{\psi}_{i}^{\mu}\Theta_{\mu\nu}(z_{0})w^{i\nu}\psi_N+h.c,
\end{equation}
\begin{equation}
{\cal L}_{\pi\Delta N}^{(2)}=\frac{i\tilde{g}_{\pi\Delta N}}{m_{N}}\bar{\psi}_{i}^{\mu}\Theta_{\mu\nu}
({z_{1}})w^{i,\nu\sigma}\gamma_{\sigma}\psi_N+h.c,
\end{equation}
 
We need all $\Delta N$ vertices is up to ${\cal O}(\epsilon^{2})$. However the following term must be included even though it is apparently third order, because it produces one ${\cal O}(\epsilon^{2})$ 
 $\pi\Delta N$ vertex in heavy baryon limit:
 
\begin{equation}
{\cal L}_{\pi\Delta N}^{(3)}=\frac{h_{\pi\Delta N}}{m_{N}^{2}}\bar{\psi}^{\mu}_{i}\Theta_{\mu\nu}
(z_{2})w^{i,\nu\sigma}D_{\sigma}\psi_{N}+h.c,
\end{equation}
 
The $\gamma \Delta N$ vertices are given by following terms: 
 
\begin{equation}
{\cal L}_{\gamma \Delta N}^{(1)}=\frac{iG_{1}}{2m_{N}}\bar{\psi}_{i}^{\mu}
\Theta_{\mu\nu}(x_{0})\gamma_{\rho}\gamma_{5}f_{+}^{i\rho\nu}\psi_N+h.c,
\end{equation}
\begin{equation}
{\cal L}_{\Delta N}^{(2)}={\cal L}_{\gamma \Delta N}^{(2,a)}+
{\cal L}_{\gamma \Delta N}^{(2,b)}
+{\cal L}^{(2,c)}_{\gamma \Delta N}+{\cal L}^{(2,d)}_{\gamma \Delta N}, 
\end{equation}
\begin{equation}
{\cal L}_{\gamma \Delta N}^{(2,a)}=\frac{-G_{2}}{(2m_{N})^{2}}\bar{\psi}^{\mu}_{i}
\Theta_{\mu\nu}(y_{0})\gamma_{5}f_{+}^{i\rho\nu}D_{\rho}\psi_N+h.c,
\end{equation}
\begin{equation}
{\cal L}_{\gamma \Delta N}^{(2,b)}=\frac{-G_{3}}{(2m_{N})^{2}}\bar{\psi}^{\mu}_{i}
\Theta_{\mu\nu}(y_{1})\gamma_{5}f_{+ ;\rho}^{i\rho\nu} \psi_N+h.c.
\end{equation}
\begin{equation}
{\cal L}_{\gamma \Delta N}^{(2,c)}=\frac{iG_{4}}{(2m_{N})^{2}}\bar{\psi}^{\mu}_{i}
\Theta_{\mu\nu}(y_{2})\sigma_{\lambda\rho}\gamma_{5}f_{+}^{i\lambda\rho ;\nu}\psi_{N}+h.c.
\end{equation}
\begin{equation}
{\cal L}_{\gamma \Delta N}^{(2,d)}=\frac{iG_{5}}{2m_{N}^{2}}\bar{\psi}^{\mu}_{i}
\Theta_{\mu\nu}(y_{3})\sigma_{\lambda\rho}\gamma_{5}f_{+}^{i\lambda\nu ;\rho}\psi_{N}+h.c.
\end{equation}

Again, we have to include the following terms which were not be considered in references such as \cite{GHKP}, because they produce ${\cal O}(\epsilon^{2})$ $\gamma \Delta N$ vertices: 

\begin{equation}
{\cal L}_{\gamma \Delta N}^{(3)}={\cal L}_{\gamma \Delta N}^{(3,a)}+{\cal L}_{\gamma \Delta N}^{(3,b)},
\end{equation}

\begin{equation}
{\cal L}_{\gamma \Delta N}^{(3,a)}=\frac{iG_{6}}{2m_{N}^{3}}\bar\psi^{\mu}_{i}\Theta_{\mu\nu}(t_{0})\gamma_{\lambda}\gamma_{5}f_{+}^{i\nu\lambda ;\rho}D_{\rho}\psi_{N}+h.c.
\end{equation} 
\begin{equation}
{\cal L}_{\gamma \Delta N}^{(3,b)}=\frac{iG_{7}}{2m_{N}^{3}}\bar\psi^{\mu}_{i}
\Theta_{\mu\nu}(t_{1})\gamma_{\lambda}\gamma_{5}f_{+}^{i\lambda\rho ;\nu}D_{\rho}\psi_{N}
+h.c.
\end{equation}

The tensor $\Theta_{\mu\nu}(z_{0})=g_{\mu\nu}+z_{0}\gamma_{\mu}\gamma_{\nu}$ was introduced by Peccei\cite{P} as the most general form 
obeys the invariance under the contact transformation. The so-called
``off-shell'' parameters, $z_{0}$ and $x_0$, $y_0$, etc. govern the couplings
of the off-shell spin-1/2 fields to external fields.  There has been
considerable controversy in finding conditions to fix these
parameters\cite{P}--\cite{BDM}.  
We find however, that at the end of our
calculation that final result is completely insensitive to the
off-shell parameters.  This is quite reasonable.  After all, as a
general rule in a consistent scheme, off-shell effects are not
observables; while they enter calculations in intermediate stages they
should be canceled in the final answers.  This is clear in the case of
$A$ from the KOS theorem \cite{KOS} and presumably applies for the
other off-shell parameters as well. In this context it is useful to
note that Tang and Ellis\cite{TE} explicitly showed such independence
for a somewhat simpler system than the one considered here. 

The next step in this approach consists of identifying the ``light''
and ``heavy'' degrees of freedom of spin-3/2 fields. The procedure is
in analogy to the case of spin-1/2 fields.  The situation becomes a
little more complicated here because of the off-shell spin-1/2
components associated with the Rarita-Schwinger field.
One can identify the ``light'' component as:
\begin{equation}
T^{i}_{\mu}=P^{+}_{v}P^{3/2}_{(33)\mu\nu}\psi_{i}^{\nu}(x)\exp(iM_{N}v\cdot x).
\end{equation}
The remaining component:
\begin{equation}
G_{\mu}^{i}=(g_{\mu\nu}-P^{+}_{v}P^{3/2}_{(33)\mu\nu})\psi_{i}^{\nu}(x)\exp(iM_{N}v\cdot x)
\end{equation}

The nucleon field is also split as ``light'' and ``heavy''components as 
in the usual HBChPT:
\begin{equation}
N(x)=P_{v}^{+}\psi_{N}\exp(iM_{N}v\cdot x),
\end{equation}
\begin{equation}
h(x)=P_{v}^{-}\psi_{N}\exp(iM_{N}v\cdot x).
\end{equation}

The general Lagrangian now take the form:
\begin{equation}
L_{N}=\bar{N}A_{N}N+(\bar{h}B_{N}N+h.c.)-\bar{h}C_{N}h,
\end{equation}
\begin{equation}
L_{\Delta N}=\bar{T}A_{\Delta N}N+\bar{G}B_{\Delta N}N+\bar{h}D_{N\Delta}T+\bar{h}C_{N \Delta}G+h.c.,
\end{equation} 
\begin{equation}
L_{\Delta}=\bar{T}A_{\Delta}T+(\bar{G}B_{\Delta}T+h.c.)-\bar{G}C_{\Delta}G.
\end{equation}
The matrix $A_{N}$, $B_{\Delta N}$..., admits a small energy scale expansion of the form:
\begin{equation}
A_{N}=A_{N}^{(1)}+A_{N}^{(2)}+A_{N}^{(3)}+.....,
\end{equation}
where $A_{N}^{(n)}$ is of order $\epsilon^{n}$. As emphasized in the
introduction, we denote by $\epsilon$ small quantities of order $p$,
like $m_{\pi}$ or the soft momenta, as well as mass difference
$\Delta=M_{\Delta}-M_{n}$. This mass difference is distinct from the
pion mass in the sense that it stays finite in the chiral limit but
vanishes in the large $N_{c}$ limit. However, in the physical 
world, $\Delta$ and $m_{\pi}$ are of the same scale---differing by only a 
factor of $\sim 2$. 
We therefore use a simultaneous expansion in both quantities.
It is only through this small scale expansion that we may obtain a 
low energy expansion of $\pi \Delta N$ system.
Such an expansion is in the spirit of large $N_{c}$ ChPT since 
$\Delta \propto \frac{1}{N_{c}}$.

The standard procedure is to
integrate the heavy components. This 
results in a relatively simple effective action:
\begin{equation}
S_{eff}=\int d^{4}x(\bar{T}\tilde{A}_{\Delta}T+\bar{N}\tilde{A}_{N}N+[\bar{T}\tilde{A}_{\Delta N}N+h.c.]),
\end{equation}
with
\begin{equation}
\tilde{A}_{\Delta}=A_{\Delta}+\gamma_{0}\tilde{D}_{N\Delta}^{\dagger}\gamma_{0}\tilde{C}_{N}^{-1}\tilde{D}_{N\Delta}+\gamma_{0}B_{\Delta}^{\dagger}\gamma_{0}C_{\Delta}^{-1}B_{\Delta}, 
\end{equation}
\begin{equation}
\tilde{A}_{N}=A_{N}+\gamma_{0}\tilde{B}^{\dagger}_{N}\gamma_{0}\tilde{C}^{-1}_{N}\tilde{B}_{N}+\gamma_{0}B^{\dagger}_{\Delta N}\gamma_{0}C_{\Delta}^{-1}B_{\Delta N}, 
\end{equation}
\begin{equation}
\tilde{A}_{\Delta N}=A_{\Delta N}+\gamma_{0}\tilde{D}^{\dagger}_{N \Delta}\gamma_{0}\tilde{C}_{N}^{-1}\tilde{B}_{N}+\gamma_{0}B^{\dagger}_{\Delta}\gamma_{0}C_{\Delta}^{-1}B_{\Delta N}.
\end{equation}
Here
\begin{equation}
\tilde{C}_{N}=C_{N}-C_{N \Delta}C_{\Delta}^{-1}\gamma_{0}C^{\dagger}_{N\Delta}\gamma_{0},
\end{equation}
\begin{equation}
\tilde{B}_{N}=B_{N}+C_{N \Delta}C_{\Delta}^{-1}B_{\Delta N},
\end{equation}
\begin{equation}
\tilde{D}_{N \Delta}=D_{N \Delta}+C_{N \Delta}C_{\Delta}^{-1}B_{\Delta}.
\end{equation}
Note $C_{\Delta}$ is a $5\times 5$ matrix\cite{HHK2}. The effect of
heavy degrees of freedom shows up at order $\epsilon^2$.  In order to
calculate a given process to order $\epsilon^{n}$, it is sufficient to
construct matrix A to the same order, $\epsilon^{n}$, B and D to order
$\epsilon^{n-1}$, and C to order $\epsilon^{n-2}$. 

\section{ Loop Graphs}

To order $\epsilon^3$, only one-loop graphs with the lowest order of vertex need be considered.
The vertices we need are from $A^{(1)}_{N}$,$A^{(1)}_{\Delta}$ and
$A^{(1)}_{\Delta N}$:
\begin{equation}
A^{(1)}_{N}=i v \cdot D+g_{A} S \cdot u ,
\end{equation}
\begin{equation}
A^{(1)}_{\Delta N}=g_{\pi\Delta N}w^{i}_{\mu} ,
\end{equation}
\begin{equation}
A^{(1)}_{\Delta}=-[iv \cdot D^{ij}-\Delta \delta^{ij}+g_{1}S \cdot u^{ij}]g_{\mu \nu} \;.
\end{equation}
Here $S_{\mu}=\frac{i}{2}\gamma_{5}\sigma_{\mu\nu}v^{\nu}$ denotes the Pauli-Lubanski spin vector.
 From $A_{N}^{(1)}$, $A_{\Delta}^{(1)}$ we determine the propagators in momentum space with residual soft momentum: $r_{\mu}=p_{\mu}-M_{N}v_{\mu}$:
\begin{equation}
S^{1/2}(v\cdot r)=\frac{i}{v\cdot r+i\epsilon},
\end{equation}
\begin{equation}
S^{3/2}_{\mu\nu}(v\cdot r)=\frac{-iP^{3/2}_{\mu\nu}}{v\cdot r-\Delta+i\epsilon}\xi^{ij}_{I=3/2},
\end{equation}
with $P_{\mu\nu}^{3/2}$ denoting the spin 3/2 projector in d-dimensions:
\begin{equation}
P_{\mu\nu}^{3/2}=g_{\mu\nu}-v_{\mu}v_{\nu}+\frac{4}{d-4}S_{\mu}S_{\nu},
\end{equation}
and 
\begin{equation}
\xi_{I=3/2}^{ij}=\delta_{ij}-\frac{1}{3}\tau^{i}\tau^{j},
\end{equation}
denotes an isospin 3/2 projector.
At present,  the coupling constant $g_{\pi \Delta N}$ 
has not been extracted from data within the context of a systematic implementation of the small scale expansion.\\ 

For simplicity, we work in the center of mass. We choose 
$v_\mu = (1,0,0,0)$ which is equivalent to working in (or close to) the center
of mass frame, and the entire calculation is done in Coulomb Gauge:
$\epsilon\cdot v$=0. This choice greatly reduces the number of graphs
which contribute. The disadvantage is that the gauge invariance is no
longer manifest.  The properties of the ``light'' components of the
delta:
$$\gamma_{\mu}T^{\mu i}(x)=v_{\mu}T^{\mu i}(x)=0,$$ and $$S_{\mu}
T^{\mu,i}(x)=0,$$ 
also ensures that many graphs vanish identically.  Note that $\pi\pi
\Delta N$ vertex, unlike $\pi\pi NN$ vertex, starts from
${\cal O}(\epsilon^{3})$, and since $\Delta$-isobars must be the intermediate
states, the $\pi\pi\Delta\Delta$ vertex and $\pi\pi\Delta N$ vertex do not 
enter our calculation. In Figs. 1, 2 all graphs are shown (1-a,b,c are 
grouped in gauge invariant classes).

\section{Born Graphs}

The Born graphs contributing to third order are shown in Figs. 3, 4 with the 
vertices given in Table 1:\\ \vspace*{-.12in}
\begin{center} 
\begin{tabular}{|c|c|c|}
\hline\hline
Vertex & Lagrangian & Remark\\
\hline
${\cal O}(\epsilon^{0})\gamma NN$ & $A_{N}^{(1)}$ &Vanishes at coulomb Gauge\\
${\cal O}(\epsilon) \pi NN$ & $A_{N}^{(1)}$ & \\
${\cal O}(\epsilon) \gamma \pi NN$ &$A_{N}^{(1)}$&Does not contribute in $\pi^{0}$ case\\
${\cal O}(\epsilon) \gamma NN$&$\gamma_{0}B^{(1)\dagger}_{N}\gamma_{0}C_{N}^{(0)-1}B_{N}^{(1)},A_{N}^{(2)}$&\\
${\cal O}(\epsilon^{2})\pi NN$ &$\gamma_{0}B^{(1)\dagger}_{N}\gamma_{0}C_{N}^{(0)-1}B_{N}^{(1)}$&\\
${\cal O}(\epsilon^{2})\gamma \pi NN$ & $\gamma_{0}B^{(1)\dagger}_{N}\gamma_{0}C_{N}^{(0)-1}B_{N}^{(1)}$&\\
${\cal O}(\epsilon^{2}) \gamma NN$&$\gamma_{0}B^{(1)\dagger}_{N}\gamma_{0}C_{N}^{(0)-1}B_{N}^{(2)},\gamma_{0}B^{(1)\dagger}_{N}\gamma_{0}C_{N}^{(1)-1}B^{(1)}$&\\
${\cal O}(\epsilon^{2})\pi\pi NN$&$A_{N}^{(1)}$&\\
${\cal O}(\epsilon^{3}) \gamma \pi NN$ &$A_{N}^{(3)},\gamma_{0}B_{N}^{(1)\dagger}\gamma_{0}C_{N}^{(0)-1}B_{N}^{(2)},\gamma_{0}B_{\Delta N}^{(2)\dagger}\gamma_{0}C_{\Delta}^{(0)-1}B_{\Delta N}^{(1)}$&\\
${\cal O}(\epsilon^{3}) \gamma \pi NN$ &$\gamma_{0}B_{N}^{(1)\dagger}\gamma_{0}C_{N}^{(1)-1}B_{N}^{(1)}$&\\
${\cal O}(\epsilon)\gamma N \Delta$&$A_{\Delta N}^{(2)}$&\\
${\cal O}(\epsilon)\pi N \Delta $&$A_{\Delta N}^{(1)}$&\\
${\cal O}(\epsilon^{2})\gamma N\Delta $&$A_{\Delta N}^{(2)},\gamma_{0}D_{N \Delta}^{(2)\dagger}\gamma_{0}C_{N}^{(0)-1}B_{N}^{(1)},\gamma_{0}B_{\Delta}^{(1)\dagger}\gamma_{0}C_{\Delta}^{(0)-1}B_{\Delta N}^{(2)}$&\\
${\cal O}(\epsilon^{2})\pi N\Delta$&$A_{\Delta N}^{(2)},\gamma_{0}B_{\Delta}^{(1)\dagger}\gamma_{0}C_{\Delta}^{(0)-1}B_{\Delta N}^{(1)}$&\\
${\cal O}(\epsilon)\pi\Delta\Delta$ &$A_{\Delta}^{(1)}$&\\
${\cal O}(\epsilon)\gamma \pi\Delta\Delta$ &$A_{\Delta}^{(1)}$&\\
\hline\hline 
\end{tabular}

\vspace*{.1in}
{\sf Table 1: Vertices for Born graphs in pion photoproduction.}\\
\end{center}

The  $\gamma N \Delta$ vertex is the focus of our study here. The leading 
order $\gamma N\Delta$ vertex is a  pure M1 one, but through 1/M expansion,
the coupling constants $G_{1}$ and $G_{2}$ still contribute to E2 
transitions. The ${\cal O}(\epsilon^{2})$ vertex with $G_{4}$ is the same with the one with $G_{5}$
and $G_{7}$, therefore we use $G_{4}$ to represent the 
$G_{4}+G_{5}+G_{7}$. 
Similar situation occurs in the $\pi\Delta N$ vertices: the ${\cal O}(\epsilon^{2})$ vertex with
 $h_{\pi\Delta N}$ is the same with $\tilde{g}_{\pi\Delta N}$, so we also use
 $\tilde{g}_{\pi\Delta N}$ to represent the $h_{\pi\Delta N}+\tilde{g}_{\pi\Delta N}$. Of course if we go to higher order calculation, these different coefficients should contribute in different way.

Note also that the heavy and off-shell spin 1/2 component of $\Delta$-isobars modify the 
NN vertices, up to third order giving nonvanishing  contributions.
The off-shell parameters show up through the 1/M expansion.  The combinations 
$h_{1}=y_{0}G_{2}-2x_{0}G_{1}-2G_{1}$, and $h_{2}=-2x_{0}G_{1}+y_{0}G_{2}$ appear in the 
${\cal O}(\epsilon^{2})\gamma \Delta N$ vertex.  However, the amplitudes 
depend  only on $h_{1}-h_{2}$, which is $2G_{1}$ and is independent of 
$x_{0}$, $y_{0}$. (Actually the vertex itself can be shown as independent of 
$x_{0}$, $y_{0}$ if the on-shell constraint: $S_{\mu}T^{\mu}_{i}=0$ is implemented, 
then our vertex is identical with the one in \cite{GHKP}.)

The off-shell parameters are also encountered in the 
$O(\epsilon^{3})\gamma \pi NN$ vertex, but they cannot be
distinguished from $b_{9}$; thus,  in our calculation, the values of 
off-shell parameters are irrelevant.
As discussed earlier, this is expected on very general grounds. 
Also the counterterm in ${\cal L}_{\pi\pi}^{(4)}$ appears in Fig. 3-C-7.

\section{ Renormalization}
The one-loop diagrams contain divergences and have to be regularized 
and renormalized. Here we will use dimension regularization.   
The unrenormalized coefficients are then related to the renormalized 
(scale dependent) ones according to
\begin{equation}
b_{i}=b^{r}_{i}(\mu)+(4\pi)^{2}\beta_{i}L \; ,
\end{equation}
\begin{equation}
L\equiv\frac{\mu^{d-4}}{(4\pi)^{2}}(\frac{1}{d-4}-\frac{1}{2}[\ln(4\pi)+1+\Gamma^{'}(1)]), 
\end{equation}
where $b_i$ represents a generic coefficient and  the $\beta_{i}^{'}$s  are the associated dimensionless 
coefficients which govern the scale dependence
of $b_{i}^{r}(\mu)$:
\begin{equation}
b_{i}^{r}(\mu)=b_{i}^{r}(\mu_{0})+\beta_{i} \ln\frac{\mu_{0}}{\mu},
\end{equation}
Ecker et al.\cite{EM} have calculated all $\beta_{i}$ in a theory  without an explicit  $\Delta$ resonance 
and obtained
\begin{equation}
\begin{array}[c]{c}
\beta_{9}=\beta_{10}=\beta_{18}=\beta_{19}=\beta_{20}=\beta_{23}=0,\\ [.12in]
\beta_{17}=\frac{g_{A}}{2}+g_{A}^{3},\, \beta_{21}=g_{A}+g_{A}^{3},\, 
\beta_{22}=-g_{A}^{3}.
\end{array}
\end{equation}

(At first glance, the $\beta$ functions of \cite{EM} and \cite{FMS}
are different but actually they are equivalent:  Some
terms which are proportional to the nucleon EOM are kept in \cite{FMS},
but transformed away in \cite{EM}. For convenience we adopt the $\beta$
functions of \cite{EM}.)

 The $\Delta-\pi$ loop correction to $\kappa_{v}$ is
: 

\begin{equation}
\begin{array}[c]{c}
\kappa_{v}(\mu)=\dot{\kappa}_{v}-\frac{m_{\pi}M_{N}g_{A}^{2}}
{4\pi F_{\pi}^{2}}+\frac{\Delta M_{N}}{(F_{\pi})^{2}}
\frac{32g_{\pi\Delta N}^{2}}{9}
(L+\frac{1}{16\pi^{2}}\ln\frac{m_{\pi}}{\mu}+
\frac{\sqrt{\Delta^{2}-m_{\pi}^{2}}}{16\pi^{2}\Delta}\ln[R])\; 
\end{array}
\end{equation}
\begin{equation}
\begin{small}
R=\frac{\Delta}{m_{\pi}}+\sqrt{(\frac{\Delta}{m_{\pi}})^{2}-1},
\end{small}
\end{equation}
 
Now, collecting all other divergences, we get following results:   
\begin{equation}
\begin{array}[c]{c}
G_{1}(\mu)+\frac{\Delta}{m_{N}}(G_{6}(\mu)-G_{4}(\mu))=
\dot{G}_{1}+\frac{\Delta}{m_{N}}(\dot{G}_{6}-\dot{G}_{4})+\frac{\Delta M_{N}}
{F_{\pi}^{2}}\frac{2g_{\pi \Delta N}}{3}
(g_{A}-\frac{35}{9}g_{1})(L+\frac{1}{16\pi^{2}}\ln\frac{m_{\pi}}{\mu})\;
\end{array}
\end{equation}
 
\begin{equation}
G_{2}(\mu)+4G_{4}(\mu)=\dot{G}_{2}+4\dot{G}_{4}+\frac{M^{2}_{N}}
{F_{\pi}^{2}}\frac{8g_{\pi \Delta N}}
{3}(g_{A}-\frac{5}{9}g_{1})(L+\frac{1}{16\pi^{2}}\ln\frac{m_{\pi}}{\mu}),
\end{equation}
 
where the $G_1$ and $G_2$ are the coefficients associated  
with the $\gamma$$\Delta$N vertex and the $\cdot$ over the $G$'s 
indicates it is taken as lowest  order value in the expansion---in 
other words, 
taken in the limit: $m_{\pi}\rightarrow 0$, $\Delta \rightarrow 0$, with 
$\frac{\Delta}{m_{\pi}}$ fixed.
The remaining  divergences are absorbed by altering the  $\beta_{i}$.
The $\pi^{0}$ photoproduction amplitude can be
renormalized only if the $\beta$ functions are modified in this way:
\begin{equation}
\beta_{9}+\frac{2}{9}\frac{(4\pi F_{\pi})^{2}}{M_{N}^{2}}g_{\pi\Delta N}
(\beta_{G6}-\beta_{G4})=\frac{2}{27}g_{\pi\Delta N}^{2}(g_{A}-\frac{5}{9}g_{1}).
\end{equation}
\begin{equation}
\beta_{10}=0.
\end{equation}
Here
\begin{equation}
G_{r}^{i}(\mu)=G_{r}^{i}(\mu_{0})+\beta_{Gi}\ln\frac{\mu_{0}}{\mu}.
\end{equation}

The charged pion amplitude also requires an alteration of
 $\beta_{21}$, $\beta_{22}$ and $\beta_{17}$  :
\begin{equation}
\beta_{22}+\frac{(4\pi F_{\pi})^{2}}{9M_{N}^{2}}g_{\pi \Delta N}(\beta_{G6}-\beta_{G4})=
-g_{A}^{3}+\frac{40}{27}g_{A}g_{\pi \Delta N}^{2}-\frac{80}{243}g_{1}g_{\pi\Delta N}^{2},
\end{equation}
\begin{equation}
\beta_{21}=g_{A}-\beta_{22},
\end{equation}
\begin{equation}
\beta_{17}=\frac{g_{A}}{2}+g_{A}^{3}-\frac{50}{81}g_{1}g_{\pi\Delta N}^{2}+\frac{2}{9}g_{A}g_{\pi\Delta N}^{2}.
\end{equation}

As Kambor pointed out\cite{k}, one cannot identify the coupling
constants of theory including delta degrees of freedom with those in
HBChPT, even if they multiply the same structures in effective
Lagrangians. The reason is the process of integrating out the
additional degrees of freedom leads to a (general infinite)
renormalization of the bare coupling constants of the underlying
theory. So if we keep $\Delta$ finite, the bare coupling constants in
our Lagrangians will differ with the usual one in HBChPT even in the
chiral limit. From (83),(84) we have:
\begin{equation}
\kappa_{v}=\kappa^{\chi}_{v}-\frac{m_{\pi}M_{N}g_{A}^{2}}{4\pi F_{\pi}^{2}}-\frac{2m_{\pi} M_{N}}{9\pi^{2}F_{\pi}^{2}}g_{\pi\Delta N}^{2}(x\ln 2x-\sqrt{x^{2}-1}\ln[x+\sqrt{x^{2}-1}]),
\end{equation}
\begin{equation}
\kappa_{v}^{\chi}=\dot{\kappa}_{v}+\frac{2\Delta M_{N}}{9\pi^{2}F_{\pi}^{2}}g_{\pi\Delta N}^{2}(16\pi^{2}L+\ln \frac{2\Delta}{\mu}),
\end{equation}
\begin{equation}
G_{1}=G_{1}^{\chi}+\frac{25m_{\pi} M_{N}}{72\pi^{2}F_{\pi}^{2}}g_{1}g_{\pi\Delta N}(x\ln 2x) ,
\end{equation}
\begin{equation}
G_{1}^{\chi}=\dot{G}_{1}-\frac{25\Delta M_{N}}{72\pi^{2}F_{\pi}^{2}}g_{1}g_{\pi\Delta N}(16\pi^{2}L+\ln \frac{2\Delta}{\mu}).
\end{equation}
Here $x\equiv \frac{\Delta}{m_{\pi}}.$
 These constants ; $m_{N}$ and $c_{1}$ are also infinitely 
renormalized:
\begin{equation}
m_{N}=m^{\chi}-4c^{\chi}_{1}M_{\pi}^{2}-\frac{3g_{A}^{2}M_{\pi}^{3}}{32\pi F_{\pi}^{2}}+\frac{g_{\pi\Delta N}^{2}M_{\pi}^{3}}{12\pi^{2}F_{\pi}^{2}}R(\frac{\Delta}{M_{\pi}}),      
\end{equation}
where
\begin{equation}
R(x)\equiv -4(x^{2}-1)^{3/2}\ln(x+\sqrt{x^{2}-1})+4x^{3}\ln 2x-6x\ln x,
\end{equation}
\begin{equation}
m^{\chi}=\dot{m}+\frac{g_{\pi \Delta N}^{2}}{3F_{\pi}^{2}}\Delta^{3}(-16 L+\frac{1}{\pi^{2}}[\frac{1}{2}-\ln\frac{2\Delta}{\mu}]),
\end{equation}
\begin{equation}
c_{1}^{\chi}=\dot{c}_{1}+\frac{g_{\pi\Delta N}^{2}}{8F_{\pi}^{2}}\Delta(-16 L+\frac{1}{\pi^{2}}[\frac{1}{3}-\ln\frac{\Delta}{\mu}]).
\end{equation}
Another interesting example is mass of isobars \cite{BM}:(Here $d_{1}$ is the coefficient associated with the counterterm : $\bar{\Delta}\Delta Tr(\chi_{+})$.)
\begin{equation}
m_{\Delta}=m^{\chi}+\Delta^{\chi}-4d^{\chi}_{1}m_{\pi}^{2}-\frac{25g_{1}^{2}M_{\pi}^{2}}{864\pi F_{\pi}^{2}}-\frac{g_{\pi\Delta N}^{2}M_{\pi}^{3}}{48\pi^{2}F_{\pi}^{2}}R(\frac{\Delta}{M_{\pi}}),
\end{equation}
\begin{equation}
\Delta^{\chi}=\Delta+\frac{g_{\pi\Delta N}^{2}}{3F_{\pi}^{2}}\Delta^{3}(
20L+\frac{5}{4\pi^{2}}\ln\frac{2\Delta}{\mu}-\frac{3}{8\pi^{2}}),
\end{equation}
\begin{equation}
d^{\chi}_{1}=\dot{d}_{1}+\frac{g_{\pi\Delta N}^{2}}{2F_{\pi}^{2}}\Delta(L+
\frac{1}{16\pi^{2}}\ln\frac{2\Delta}{\mu}-\frac{1}{8\pi^{2}}),
\end{equation}
 The renormalization of the  axial coupling constant is 
similar but  more complicated\cite{BFHM}:
\begin{equation} 
g_{A}=g_{A}^{\chi}+\frac{4m_{\pi}^{2}}{(4\pi F_{\pi})^{2}}\{(b_{17}^{r}-\frac{b_{19}^{r}}{2})+S(\frac{\Delta}{M_{\pi}})\},
\end{equation}
\begin{equation}
g_{A}^{\chi}=
\dot{g}_{A}+\frac{4g_{A}g_{\pi\Delta N}^{2}}{27\pi^{2}F_{\pi}^{2}}\Delta^{2}\{16\pi^{2} L+\ln\frac{2\Delta}{\mu}+\frac{1}{2}\},
\end{equation}
where
\begin{equation}
\begin{array}[c]{c}
S(x)\equiv a_{1}+a_{2}x^{-1}+a_{3}x^{2}\ln 2x 
+a_{4}x\sqrt{x^{2}-1}\ln [x+\sqrt{x^{2}-1}]\\[.09in]
+a_{5}\frac{(x^{2}-1)^{3/2}}{x} \ln [x+\sqrt{x^{2}-1}], 
\end{array}
\end{equation}
\begin{equation}
a_{1}=\frac{-4}{3}g_{A}^{3}-\frac{136}{27}g_{A}g_{\pi\Delta N}^{2}+\frac{200}{243}g_{1}g_{\pi\Delta N}^{2},
\end{equation}
\begin{equation}
a_{2}=\frac{64\pi}{27}g_{A}g_{\pi\Delta N}^{2},
\end{equation}
\begin{equation}
a_{3}=-\frac{304}{27}g_{A}g_{\pi\Delta N}^{2}+\frac{400}{81}g_{1}g_{\pi\Delta N}^{2},
\end{equation}
\begin{equation}
a_{4}=16g_{A}g_{\pi\Delta N}^{2}-\frac{400}{81}g_{1}g_{\pi\Delta N}^{2},
\end{equation}
\begin{equation}
a_{5}=-\frac{128}{27}g_{A}g_{\pi\Delta N}^{2}.
\end{equation}

In general, the LEC's in the present expansion are different from HBChPT without an explicit $\Delta$ degree of freedom. 
The reason for this is obvious: processes including the explicit $\Delta$ contribute and serve to renormalize the parameters.
 However, in the chiral limit of $m_\pi =0$,  $\kappa_{v}^{\chi}$, $G_{1}^{\chi}$, $m^{\chi}$, $c_{1}^{\chi}$ and $g_{A}^{\chi}$ can be identified with
the analogous  LECs in HBChPT without $\Delta$'s.   
In contrast, parameters of $\dot{\kappa}_{v}$, $\dot{G}_{1}$, $\dot{g}_{A}$, $\dot{m}_{N}$ and $\dot{c}_{1}$ 
are defined in the limit: $\Delta\rightarrow 0$, $m_{\pi}\rightarrow 0$, $\frac{\Delta}{m_{\pi}}$ fixed.

There is an interesting formal limit to consider apart from the chiral limit;
namely the large $N_c$ limit. 
 Recall in this
limit one naturally has
$\Delta \rightarrow 0$ with $m_{\pi}$ finite.
Moreover, in the large $N_c$ limit, $g_{A}=\frac{5}{9}g_{1}=\frac{2\sqrt{2}}{3}g_{\pi\Delta N}.$ 
Inserting this into our expressions for the $\beta$'s we find:
\begin{equation}
\beta_{17}=\frac{g_{A}}{2}.
\end{equation}
Actually such a result is required by the consistency of $N_{c}$ counting rules. Note
that $g_{A}$ is ${\cal O}(N_{c})$ quantity, and $F_{\pi} \sim {\cal
O}(\sqrt{N_{c}})$ and $m_{\pi} \sim {\cal O}(1)$; therefore all terms
with $g_{A}^{3}$ must vanish in the large $N_{c}$ limit.  Such a
cancelation is possible only when the $\Delta$-isobar degrees of
freedom are included \cite{CB},\cite{TDC}.

\section {Comparison with Experiment}
  The values of unknown parameters such as $G_{1},G_{2}...$ 
are expected to be extracted from the experimental data, and 
to be used to make predictions in other processes. However, as mentioned in the Introduction, there is an obstacle:  
 The amplitudes
at the $\Delta$ pole  diverge, and amplitudes in the vicinity of the pole clearly cannot be taken seriously.

At first glance, this appears to be a  fatal problem for this approach since our interest is precisely in the $\Delta$ region. 
One natural way to cure this is to use a dressed delta propagator, i.e, put the self-energy part $\Sigma(E)$ in the
 delta propagator. With a self-energy included, amplitudes become smooth since the imaginary part in $\Sigma(E)$ shifts the pole from the real axis. However this approach breaks the
power counting scheme in its purest form because it requires that part of the interaction (those associated with $\Delta$
 decay into
pion plus nucleon) be iterated to all orders while other parts are not.   
The overall expansion becomes questionable, and as we pointed out in the previous section, the power counting scheme is our only way to make predictions consistently.

At a technical level, the problem is that away from the $\Delta$ pole, the $\Delta$ propagator
 is ${\cal O}(\epsilon^{-1})$, while in the immediate vicinity the pole is ${\cal O}(\epsilon^{-2})$.  As one moves up the resonance,
 the behavior changes, making a systematic treatment problematic.
One purely phenomenological alternative is to simply insert the empirical decay width $\Gamma$ in the invariant amplitude as Adelseck et al. have done\cite{ABE}.  Such a scheme is not systematic, however; and moreover, unitarity is violated\cite{BDM}.

Therefore we suggest that instead of directly comparing the theoretically 
calculated amplitudes with the  experimentally extracted 
ones at all energies, we only compare weighted integrals of the amplitudes.
For example, 
\begin{equation}
\bar{{\cal M}}^{(n)}_{l\pm}=\int^{E_{max}}_{m_{\pi}}{\cal M}_{l\pm}(E)W_{n}(E)dE .   
\end{equation}
Here $E$ is the energy of the photon in the c.m frame, $W_{n}(E)$ is a smooth weight function, and n is an integer index specifying the particular choice of weight function.

To justify such an approach, the weight functions need to be chosen with
some care.
First of all, the hierarchy introduced by power counting must be maintained after integration. In other words, the amplitude with factor
 $\frac{1}{E-\Delta+i\epsilon}$ should not be enhanced beyond what is permitted in our power counting. To satisfy this requirement we must integrate through the entire $\Delta$ region. 
Actually there's another reason to do so:  
Recall that in the vicinity of the pole, the theoretical  resonant amplitude
has a $\delta$ function in the strength
 through $$\lim_{\epsilon\rightarrow 0}\frac{1}{E-\Delta+i\epsilon}=P(\frac{1}{E-\Delta})-i\pi\delta(E-\Delta),$$
but the actual resonant amplitudes will be more like a Breit-Wigner form
and one expects  the imaginary part to spread the 
width $\Gamma $. Therefore, the weight functions are required to  
cover the whole $\Delta$ region to keep the full information of the 
imaginary part in an experimental resonant amplitude.
On the other hand, $E_{max}$ could not be put too high. The ${\cal O}(\epsilon^{4})$ contribution of amplitudes becomes more important when energy increases and our calculation loses its predictive power in the higher energy region. To satisfy both 
requirements, the best place
 for $E_{max}$ 
is at the upper end of the $\Delta$ region.  Since we are interested in power counting it is sensible to
look at a Taylor series for the $W_n(E)$:
\begin{equation}
W_{n}(E)=a^{(0)}_{n}+a^{(1)}_{n}(\frac{E}{m_{p}})+a^{(2)}_{n}
(\frac{E}{m_{p}})^{2}+a^{(3)}_{n}(\frac{E}{m_{p}})^{3}+
a^{(4)}_{n}(\frac{E}{m_{p}})^{4}+...
\end{equation}

It is clear that the effect of higher order coefficients such as
$a^{(4)}_{n}$ do not contain reliable information since they can not
be distinguished from the effect of ${\cal O}(\epsilon^{4})$
amplitudes which are absent from our calculation. Accordingly we can
choose
$$1,(\frac{E}{m_{p}}),
(\frac{E}{m_{p}})^{2},(\frac{E}{m_{p}})^{3},$$
as our ``basis functions''. Any $W_{n}(E)$ is equivalent to their
linear combination once the higher order terms are thrown away.  Thus
we choose $W_{n}(E)$ as
$$W_{n}(E)=(\frac{E}{m_{p}})^{n}.$$
The preceding analysis also suggests another advantage of studying these integrated quantities apart from the problem of the $\Delta$ pole.
Note there is a finite amount of information which can be extracted from a systematic expansion
at a finite order.  Direct predictions for cross sections as functions of energy formally have
an infinite information content since it takes an infinite number of parameters to describe an 
arbitrary function.  Clearly, much of the information content contained in a predicted functional 
dependence has considerable correlations.  It is useful, therefore, to construct a number of
 discrete observables, such as our integrals which characterize the energy dependences.  The scheme
proposed here corresponds to picking the maximum number of independent predictions which we
can make at this order.

We  integrate from the threshold through the entire $\Delta$ region:
\begin{equation}
\bar{{\cal M}}^{(n)}_{l\pm}=\frac{1}{m_{p}}
\int^{E_{max}}_{m_{\pi}}{\cal M}_{l\pm}(E)
(\frac{E}{m_{p}})^{n}d\omega .
\end{equation}
The additional factor $1/m_{p}$ is only to ensure that $\bar{{\cal M}}$ 
has the same dimension as ${\cal M}$.
The values of the unknown parameters can be fit from the amplitudes as 
follows:
\begin{equation}
\begin{array}[c]{c}
Re\bar{P}^{\pi^{0}P}_{i}=eg_{A}\zeta^{A}_{i}+eg_{A}\dot{\kappa}_{p}
\zeta^{B}_{i;B}+eg_{A}\kappa_{p}\zeta_{i;R}^{B}
+eg_{A}(1+\dot{\kappa}_{p})\tilde{c}_{1}\zeta^{C}_{i}\\ [.09in]
+eg_{A}^{3}\zeta^{D}_{i}+eg_{\pi \Delta N}\dot{G}_{1}\zeta^{E}_{i;B}
+eg_{\pi\Delta N}G_{1}\zeta_{i;R}^{E}
+eg_{\pi \Delta N}\tilde{G}_{2}\zeta^{F}_{i}+eg_{\pi\Delta N}\tilde{G}_{6}\zeta^{G}_{i}\\[.09in]
+e\tilde{g}_{\pi\Delta N}\dot{G}_{1}\zeta^{H}_{i}+eg_{\pi \Delta N}^{2}g_{A}\zeta^{K}_{i}+eg_{\pi \Delta N}^{2}g_{1}\zeta^{L}_{i}+e\tilde{b}_{9}\zeta^{M}_{i}
,\\[.09in]
i=1,2,3.
\end{array}
\end{equation}
\begin{equation}
\begin{array}[c]{c}
\frac{1}{\pi}Im\bar{P}^{\pi^{0}P}_{i}=eg_{A}^{3}\xi^{D}_{i}+eg_{\pi\Delta N}\dot{G}_{1}\xi^{E}_{i;B}+eg_{\pi\Delta N}G_{1}\xi^{E}_{i;R}+eg_{\pi\Delta N}\tilde{G}_{2}\xi^{F}_{i}+eg_{\pi\Delta N}\tilde{G}_{6}\xi^{G}_{i}\\[.09in]
+e\tilde{g}_{\pi\Delta N}\dot{G}_{1}\xi^{H}_{i}+eg_{\pi \Delta N}^{2}g_{A}\xi^{K}_{i}+eg_{\pi \Delta N}^{2}g_{1}\xi^{L}_{i},\\[.09in]
i=1,2,3.
\end{array}
\end{equation}

Here
$\tilde{G}_{2}=G_{2}+4G_{4}$, $\tilde{G}_{6}=G_{6}-G_{4}$,  $\tilde{c}_{1}=m_{N}\dot{c}_{1}$, $\kappa_{p}=\frac{1}{2}(\kappa_{v}+\kappa_{s}), \tilde{b}_{9}=b_{9}-b_{10}-\frac{(4\pi F_{\pi})^{2}}{6m_{N}^{2}}g_{\pi \Delta N}\dot{G}_{1}(1+4x+4z+12xz).$ $\dot{\kappa}_{p}$ means the parameter taken in the limit:$\Delta \rightarrow 0$
, $m_{\pi} \rightarrow 0$, $\frac{\Delta}{m_{\pi}}$ fixed, and $\kappa_{p}$; 
 the $\pi$ loop correction is included in this.
 
The first four terms in (146) are from tree graphs without delta; the 
sixth to eleventh terms are due to tree graphs with delta. Note that such  
tree graphs also contribute to the imaginary parts of amplitudes due 
to the delta function in $\frac{1}{E-\Delta+i\epsilon}$. The fifth 
term is from loop graphs without delta; the twelfth and thirteenth terms
are $\Delta-\pi$ loop contributions; the last term, which only 
appears in $P_{3}$, is due to the counterterms in ${\cal L}^{(3)}_{\pi NN}$.  
Note that the quantities (as they must be),  
like $\xi^{K}_{i},\xi^{L}_{i}$ are $\mu$-dependent, however final 
amplitudes are independent of $\mu$ because the $\kappa_{v}$, $G_{1}$, $\tilde{G}_{2}$ 
 and $\tilde{G}_{6}$ are also $\mu$-dependent, and compensate the ones from the loop.
Note that the contribution of $\tilde{G}_{6}$ always couples with the one of $\tilde{g}_{\pi\Delta N}$, it's impossible to separate them by fitting the pion photoproduction data. But only the latter appears in the $\pi-N$ scattering amplitudes, therefore their values can be determined by both experimental data.

It is obvious that power counting is preserved in this scheme with this set of weight functions. The ${\cal O}(\epsilon^{2})$ 
amplitudes represented by $\zeta_{i}^{A}$, $\zeta^{B}_{i;R}$ or $\zeta^{E}_{i;R}$, (i=1,2) are significantly  larger than the other terms, and  $\zeta^{K}_{1;\pi}$ or $\zeta^{L}_{2;\pi}$ and $\zeta^{F}_{i}$, $\zeta_{i}^{G}$ are not particularly 
enhanced which  shows that the
wild behavior of $\frac{1}{E-\Delta+i\epsilon}$ is tamed by our
weight functions.

We set $\Delta$=294 Mev, $F_{\pi}$=92.4 Mev, $M_{N}$=938 Mev, 
$E_{max}$=340 Mev, and $\mu$=500 Mev. The following results are given (in the unit of $10^{-4}/m_{\pi}$):
\newpage
\begin{center}
\begin{flushleft}
\vspace{4. mm}
\begin{small}
\begin{tabular}{|c|c|c|c|c|c|c|c|c|c|c|c|c|c|c|}
\hline\hline
n&$\zeta^{A}_{1}$&$\zeta^{B}_{1;R}$&$\zeta^{B}_{1;B}$&$\zeta^{C}_{1}$&$\zeta_{1}^{D}$&$\zeta_{1;R}^{E}$&$\zeta_{1;B}^{E}$&$\zeta_{1}^{F}$&$\zeta_{1}^{G}$&$\zeta^{H}_{1}$&$\zeta_{1;\pi}^{K}$&$\zeta_{1;\Delta}^{K}$&$\zeta_{1;\pi}^{L}$&$\zeta_{1;\Delta}^{L}$\\
\hline
0&23.89&26.93&-1.04&2.40.&9.23&-12.07&-3.92&0.74&-2.96&-2.96&-1.07&15.62&6.75&0\\
\hline
1&6.60&7.52&-0.36&0.67&2.66&-4.65&1.03&0.29&-1.18&-1.18&-0.72&4.31&3.04&0\\
\hline
2&1.90&2.19&-0.12&0.20&0.80&-1.67&-0.28&0.11&-0.44&-0.44&-0.31&1.18&1.17&0\\
\hline
3&0.57&0.66&-0.04&0.06&0.24&-0.58&-0.08&0.04&-0.16&-0.16&-0.12&0.32&0.42&0\\
\hline
\end{tabular}
\end{small}
\noindent
\vspace{4. mm}
\begin{tabular}{|c|c|c|c|c|c|c|c|c|c|c|}
\hline
n&$\xi^{D}_{1}$&$\xi^{E}_{1;R}$&$\xi^{E}_{1;B}$&$\xi_{1}^{F}$&$\xi^{G}_{1}$&$\xi^{H}_{1}$&$\xi^{K}_{1;\pi}$&$\xi_{1;\Delta}^{K}$&$\xi_{1;\pi}^{L}$&$\xi_{1;\Delta}^{L}$\\
\hline
0&-3.47&6.86&-0.95&-0.47&1.90&1.90&2.51&2.53&0&-6.53\\
\hline
1&-0.97&1.90&-0.26&-0.13&0.52&0.52&0.98&0.70&0&-1.80\\
\hline
2&-0.28&0.52&-0.07&-0.03&0.14&0.14&0.35&0.19&0&-0.50\\
\hline
3&-0.08&0.14&-0.02&-0.01&0.04&0.04&0.12&0.05&0&-0.14\\
\hline
\end{tabular}
\noindent
\vspace{4. mm}
\begin{small}
\begin{tabular}{|c|c|c|c|c|c|c|c|c|c|c|c|c|c|c|}
\hline
n&$\zeta_{2}^{A}$&$\zeta_{2;R}^{B}$&$\zeta_{2:B}^{B}$&$\zeta_{2}^{C}$&$\zeta_{2}^{D}$&$\zeta_{2;R}^{E}$&$\zeta_{2;B}^{E}$&$\zeta_{2}^{F}$&$\zeta_{2}^{G}$&$\zeta^{H}_{2}$&$\zeta^{K}_{2;\pi}$&$\zeta_{2;\Delta}^{K}$&$\zeta_{2;\pi}^{L}$&$\zeta_{2;\Delta}^{L}$\\
\hline
0&-41.53&-24.82&1.66&-2.40&-12.10&12.01&5.48&1.49&2.96&2.96&0.54&5.82&-5.36&0\\
\hline
1&-11.41&-6.91&0.54&-0.67&-3.46&4.65&1.93&0.59&1.18&1.18&-0.31&1.61&-1.94&0\\
\hline
2&-3.27&-2.00&0.18&-0.20&-1.02&1.67&0.66&0.21&0.44&0.44&-0.21&0.44&-0.67&0\\
\hline
3&-0.97&-0.60&0.06&-0.06&-0.31&0.58&0.22&0.07&0.16&0.16&-0.09&0.12&-0.22&0\\
\hline
\end{tabular}
\end{small}
\noindent
\vspace{4. mm}
\begin{tabular}{|c|c|c|c|c|c|c|c|c|c|c|}
\hline
n&$\xi_{2}^{D}$&$\xi_{2;R}^{E}$&$\xi_{2;B}^{E}$&$\xi_{2}^{F}$&$\xi_{2}^{G}$&$\xi^{H}_{2}$&$\xi^{K}_{2;\pi}$&$\xi_{2;\Delta}^{K}$&$\xi_{2;\pi}^{L}$&$\xi_{2;\Delta}^{L}$\\
\hline
0&0.80&-6.86&-1.89&-0.95&-1.90&-1.90&0.07&2.17&0&2.56\\
\hline
1&0.23&-1.90&-0.52&-0.26&-0.52&-0.52&0.12&0.70&0&0.71\\
\hline
2&0.07&-0.52&-0.14&-0.07&-0.14&-0.14&0.06&0.17&0&0.19\\
\hline
3&0.02&-0.14&-0.04&-0.02&-0.04&-0.04&0.02&0.05&0&0.05\\
\hline
\end{tabular}
\noindent
\vspace{4. mm}
\begin{small}
\begin{tabular}{|c|c|c|c|c|c|c|c|c|c|c|c|c|c|c|}
\hline
n&$\zeta_{3}^{A}$&$\zeta_{3;R}^{B}$&$\zeta_{3;B}^{B}$&$\zeta_{3}^{C}$&$\zeta_{3;R}^{E}$&$\zeta_{3;B}^{E}$&$\zeta_{3}^{F}$&$\zeta_{3}^{G}$&$\zeta^{H}_{3}$&$\zeta_{3;\pi}^{K}$&$\zeta_{3;\Delta}^{K}$&$\zeta_{3;\pi}^{L}$&$\zeta_{3;\Delta}^{L}$&$\zeta_{3}^{M}$\\
\hline
0&-0.36&3.68&-7.50&-9.32&12.41&-1.88&1.16&9.28&9.28&-0.67&9.80&4.69&0&19.26\\
\hline
1&-0.16&1.06&-2.23&-2.50&5.94&-0.41&0.41&3.28&3.28&-0.19&2.70&2.90&0&5.59\\
\hline
2&-0.06&0.32&-0.68&-0.70&2.35&-0.09&0.15&1.20&1.20&-0.05&0.75&1.24&0&1.68\\
\hline
3&-0.02&0.10&-0.21&-0.20&0.86&-0.02&0.04&0.16&0.16&0.02&0.22&0.47&0&0.52\\
\hline
\end{tabular}
\end{small}
\noindent
\vspace{4. mm}
\begin{tabular}{|c|c|c|c|c|c|c|c|c|c|}
\hline
n&$\xi_{3;R}^{E}$&$\xi_{3:B}^{E}$&$\xi_{3}^{F}$&$\xi_{3}^{G}$&$\xi_{3}^{H}$&$\xi_{3;\pi}^{K}$&$\xi_{3;\Delta}^{K}$&$\xi_{3;\pi}^{L}$&$\xi_{3;\Delta}^{L}$\\
\hline
0&-13.72&-0.95&-0.47&-3.80&-3.80&0.90&0.36&0&-9.09\\
\hline
1&-3.79&-0.26&-0.13&-1.04&-1.04&0.42&0.10&0&-2.50\\
\hline
2&-1.04&-0.07&-0.04&-0.29&-0.29&0.17&0.03&0&-0.69\\
\hline
3&-0.29&-0.02&-0.01&-0.08&-0.08&0.06&0.10&0&-0.19\\
\hline\hline
\end{tabular}
\noindent
\end{flushleft}
\end{center}

\vspace{5. mm}
\noindent {\sf Table 2: $\xi^{K}_{i;\Delta}$, $\xi^{E}_{i}$, $\xi^{G}_{i}$ and
$\xi_{i;\Delta}^{L}$ represent the contributions due to the delta
function of $\frac{1}{E-\Delta+i\epsilon}$ in the loop diagrams
like (2-A-1). The other $\xi_{i}$'s are from $\pi$ loops.}

Unfortunately, at present there is not enough data to test this theory.
This scheme only predicts integrals of amplitudes through the Delta region. 
In order to compare 
with experiment it is essential that the amplitudes be separated out so that these integrals can 
be estimated.  To do this,
 more precise and complete
measures of spin observables are required.  The cross section plus three single-spin observables determine the magnitudes of the 
amplitudes, but double-spin observables determine their relative phase.
 By carefully selecting four of the double-spin observables, one can extract all of the requisite phase without discrete 
ambiguities \cite{BDS,\cite{CT}},\cite{ALL},\cite{KeW}.
Once the experimental data of amplitudes are available, the values of these unknown parameters
 can be extracted and we can test our predictions.

A fundamental  issue is the predictive power of our method of comparison.
The scheme is designed to separate out the maximum number of independent quantities extractable
from experiment at a given in the small scale expansion.  To be predictive, there must be more observables than free parameters.
 In neutral pion photoproduction, the P-wave amplitudes are determined by 
eleven unknown parameters. With these eleven parameters 
 we have 12 integrated observables to fit. Fortunately from the 
S-wave amplitudes:
\begin{equation}
E_{0+}^{\pi^{0}p}(\omega)=E_{0+}^{Born,N}+E_{0+}^{N-\pi Loop}+E_{0+}^{Born,\Delta}.
\end{equation}
\begin{equation}
E_{0+}^{Born,N}=-\frac{eg_{A}}{24\pi 
 m_{N}F_{\pi}}\{\frac{m_{\pi}^{2}}{\omega}
+2\omega \}+\frac{eg_{A}}{48\pi 
 m_{N}^{2}F_{\pi}}\{6\omega^{2}+4m_{\pi}^{2}-
\frac{m_{\pi}^{4}}{\omega^{2}}(1-8\tilde{c}_{1})+\dot{\kappa}_{p}(4\omega^{2}-m_{\pi}^{2})\}.
\end{equation}
\begin{equation}
E_{0+}^{Loop}=\frac{eg_{A}}{64\pi^{2}F_{\pi}^{3}}[m_{\pi}^{2}\arcsin\frac{\omega}{m_{\pi}}-\omega\sqrt{m_{\pi}^{2}-\omega^{2}}].
\end{equation}
\begin{equation}
E_{0+}^{Born,\Delta}=\frac{eg_{\pi\Delta
 N}\dot{G}_{1}}{4\pi m_{N}^{2}F_{\pi}}
\frac{\omega}{\omega+\Delta}[ \frac{5\omega^{2}-2m_{\pi}^{2}}{27}+\frac{\omega(\omega^{2}-m_{\pi}^{2})}{9(\omega+\Delta)} ].
\end{equation}
here $\omega$ is the pion energy in the c.m frame.
 Four parameters: $g_{A}$, $\dot{\kappa}_{p}$, $\tilde{c}_{1}$ and 
$g_{\pi\Delta N}\dot{G}_{1}$ can be extracted without integrals because that our
 calculated $E_{0+}^{\pi^{0}p}$ doesn't diverge 
when $\omega\rightarrow \Delta$,
 therefore only 7 unknown parameters need to be fit from P-wave multipoles.
 Toward the charge pion photoproduction, the P-wave has only one new parameters: $b_{22}+b_{23}$, and again theoretical results of $E_{0+}^{\pi^{+}n}$
 never diverge in $\Delta$ region therefore
 four parameters: $g^{r}_{A}$, 
$b_{17}-\frac{b_{19}}{2}$, $b_{21}$ and $\dot{\kappa}_{n}$ can be determined
by matching them with experimental data.

Furthermore if we generalize to the case
of electroproduction, there are only two additional parameters which  need to be fit, and the 
number of observables increases since there are additional C2 amplitudes. Therefore up to ${\cal O}(\epsilon^{3})$  the predictive power of our approach is clear. If one works at higher order,
the number of basis functions increases, therefore a larger number of independent  observables becomes
 available.  However  many new parameters will be involved.   Whether one ultimately has increased the predictive power has not yet been settled.

 In conclusion, the ``small scale expansion'' provides us a systematic way to calculate the processes of the $\pi \Delta N$ system, and with the power counting scheme we can
fit the unknown parameters to make predictions.   Our method is designed to isolate the maximum number of independent predictions which can be made at a given order.

\vspace{5.00 mm}
\noindent
{\bf Acknowledgments}\\
The authors wish to thank Barry Holstein and Ulf-G Meissner for helpful discussions. The  support of  U.S. Department of Energy under grant DOE-FG02-93ER-40762 is gratefully acknowledged.
\vspace{20. mm}
\pagebreak

\section*{Appendix A:  Feynman Rules}

Here we collect the Feynman rules which are needed to calculate tree and loop
diagrams. The following notations are used:\\
l: momentum of a pion or nucleon or delta propagator;\\
k: momentum of photon;\\
q: momentum of external pion;\\
$\epsilon$: Photon polarization vector;\\
p: momentum of a nucleon in heavy mass formulation;\\
r: momentum of a delta in heavy mass formulation.\\
Isospin indices of pion are a, b, c, d..., the isospin indices of 
$\Delta$ are i, j, k..., and the spin indices of $\Delta$ are 
$\mu,\nu,\sigma$. $v_{\mu}$ is the nucleon four-velocity and 
$S_{\mu}$ is a  covariant spin-vector. We also give the 
orientation of momenta at the vertices, i.e, which are ``in''-going 
or``out''-going.  Q$\equiv$diag(1,0). Here we only present the 
ones related to $\Delta$ since others can be found in \cite{BKM2}.

\noindent
{\bf Vertices from ${\cal L}^{(1)}_{\Delta}$:}\\ \vspace*{.25in}
(1)~$\Delta$ propagator: $$\frac{-iP_{\mu\nu}^{3/2}\xi^{ij}_{I=3/2}}
{v\cdot l-\Delta+i\epsilon},$$
(2)~One pion (q out): $$\frac{-g_{1}}{F_{\pi}}\xi^{ik}_{I=3/2}
\tau^{c}\xi^{kj}_{I=3/2}(S\cdot q)g_{\mu\nu}$$
\noindent
(3)~One pion, one photon: $$\frac{-ieg_{1}}{F_{\pi}}\epsilon^{c3b}\xi^{ik}_{I=3/2}\tau^{b}\xi^{kj}_{I=3/2}(S\cdot \epsilon)g_{\mu\nu}$$
{\bf Vertices from ${\cal L}^{(2)}_{\Delta}$:}\\ \vspace*{.25in}
(1)~$\Delta$ propagator:$$\frac{-iP_{\mu\nu}^{3/2}\xi^{ij}_{I=3/2}}{(v\cdot l -\Delta+i\epsilon)^{2}}\frac{-1}{2m_{N}}(l^{2}-(v\cdot l)^{2})$$
{\bf Vertices from ${\cal L}^{(1)}_{\Delta N}$:}\\  \vspace*{.25in}
(1)~One pion (q,out):$$\frac{g_{\pi \Delta N}}{F_{\pi}}q_{\mu}\delta^{ci}$$
(2)~One photon (k,in):$$\frac{ieG_{1}}{m_{N}}\delta^{i3}[(S\cdot k)\epsilon_{\mu}-(S\cdot \epsilon)k_{\mu}]$$
(3)~One pion, one photon:$$\frac{ieg_{\pi \Delta N}}{F_{\pi}}\epsilon_{\mu}\epsilon^{c3i}$$
\noindent
{\bf Vertices from ${\cal L}^{(2)}_{\Delta N}$:}\\ \vspace*{.25in}
(1)~One pion (p in; r, q out):$$-\frac{g_{\pi \Delta N}}{m_{N}F_{\pi}}r_{\mu}
(v\cdot q)\delta^{ci}-\frac{\tilde{g}_{\pi\Delta N}}
{m_{N}F_{\pi}}q_{\mu}(v\cdot q)\delta^{ci}$$\\
(2)~One photon (k,p in; r out):$$\frac{ie}{2m_{N}^{2}}
(v\cdot k)\delta^{i3}[-h_{1}(S \cdot \epsilon)r_{\mu}+
\frac{h_{2}}{2}(S\cdot r)\epsilon_{\mu}-ih_{2}
(S\cdot r)\epsilon^{\mu\rho\alpha\beta}\epsilon_{\rho}v_{\alpha}S_{\beta}
+\frac{G_{2}}{2}\epsilon_{\mu}(S\cdot r-p)-G_{1}\epsilon_{\mu}(S\cdot r+p)]$$
$$+\frac{ie}{2m_{N}^{2}}(\epsilon\cdot v)\delta^{i3}
[h_{1}(S \cdot k)r_{\mu}-\frac{h_{2}}{2}(S \cdot r)k_{\mu}+ih_{2}
(S \cdot r)\epsilon^{\mu\rho\alpha\beta}k_{\rho}v_{\alpha}S_{\beta}-
\frac{G_{2}}{2}k_{\mu}(S \cdot r-p)+G_{1}k_{\mu}(S \cdot r+p)]$$
$$+\frac{ie}{m_{N}^{2}}(v\cdot k)\delta^{i3}[G_{6}(S\cdot k)\epsilon_{\mu}+(G_{4}-G_{6})(S\cdot \epsilon
)k_{\mu}]$$
$$+\frac{ie}{m_{N}^{2}}(\epsilon \cdot v)\delta^{i3}[-G_{6}(S\cdot \epsilon)k_{\mu}-(G_{4}-G_{6})(S\cdot k)\epsilon_{\mu}]$$,
$$\,\,\,\, h_{1}=-2G_{1}-2x_{0}G_{1}+y_{0}G_{2};\,\, h_{2}=y_{0}G_{2}-2x_{0}G_{1}.$$\\
(3)~One pion, one photon (p,k in; r,q out): $$-\frac{eg_{\pi\Delta N}}{m_{N}F_{\pi}}(Q\delta^{ci}-i\epsilon^{ic3})(v\cdot q)\epsilon_{\mu}-\frac{ieg_{\pi\Delta N}}{m_{N}F_{\pi}}r_{\mu}(\epsilon\cdot v)\epsilon^{c3i}$$
$$-\frac{i\tilde{g}_{\pi\Delta N}}{m_{N}F_{\pi}}\epsilon^{c3i}(q_{\mu}(\epsilon\cdot v)+\epsilon_{\mu}(v\cdot q)).$$

\end{document}